\documentclass[twocolumn,secnumarabic, amssymb, amsmath,tightenlines,nofootinbib,showpacs,superscriptaddress,preprintnumbers, aps,prl]{revtex4}
\usepackage{amsfonts}
\usepackage{docs}%
\usepackage{bm}%
\usepackage[colorlinks=true,linkcolor=blue,pagecolor=blue,filecolor=blue,menucolor=yellow,urlcolor=blue,citecolor=blue,anchorcolor=blue]{hyperref}%
\usepackage{graphicx}
\usepackage{dcolumn}
\usepackage{times}
\usepackage{subfigure}
 \usepackage{epsfig} 
 \usepackage{bm}
 \DeclareMathOperator{\sgn}{sgn}
 \usepackage{textcomp}
\begin{document}
\title{Exotic leaky wave radiation from anisotropic epsilon near zero metamaterials}
\author{Klaus Halterman}
\author{Simin Feng}
\affiliation{Naval Air Warface Center, Michelson Laboratory,
Physics Division,
China Lake, California 93555, USA}
\author{Viet Cuong Nguyen}
\affiliation{Photonics Research Centre, School of Electrical and Electronics Engineering,
Nanyang Technological University, 50 Nanyang Avenue, Singapore 639798}
\date{\today}


\begin{abstract}
We investigate the emission of electromagnetic
waves from  biaxial
subwavelength  metamaterials. For tunable anisotropic structures
that exhibit a vanishing
dielectric response along a given axis,
we find remarkable variation
in the launch angles of energy
associated with
the emission of leaky wave radiation.
We write closed form expressions for the energy transport velocity
and corresponding radiation angle $\varphi$, defining 
the cone of radiation emission,
both as a functions of frequency, 
and
material and geometrical parameters.  Full wave simulations exemplify the
broad range of directivity that can be achieved in these structures.
\end{abstract}

\pacs{81.05.Xj,42.25.Bs,42.82.Et
}

\maketitle
Metamaterials are composite structures engineered
with subwavelength components, and
with the purpose 
of manipulating and
directing electromagnetic (EM) radiation.
Recently
many practical applications have emerged,
and structures fabricated 
related to   
cloaking, metamaterial perfect absorbers \cite{liu2}, 
and chirality \cite{zhang2,menzel}.
For metamaterials,
the desired EM response
to the incident electric (${\bm E}$) and magnetic (${\bm H}$) fields, 
typically involves tuning the permittivity, $\epsilon$,
and permeability, $\mu$ in rather extraordinary ways. 
This includes
double negative index media (negative real parts of both $\epsilon$ and $\mu$), 
single negative index media (negative real part of $\epsilon$ {\it or} $\mu$), 
matched impedance zero index media \cite{zio,cuong} 
(real part of $\epsilon$ and $\mu$ is 
near zero),
and
epsilon near zero (ENZ) media (real part of $\epsilon$ is near zero).
Scenarios involving ENZ media in particular have gained prominence lately
as useful components to radiative systems
over a broad range of the EM spectrum \cite{enoch, eng, alu}.

In conjunction with ENZ developments, 
there have also been 
advances in infrared metamaterials, 
where thermal emitters \cite{laroche}, optical switches \cite{shen},
and negative index metamaterials \cite{zhang,zhang2} have been fabricated.
Due to the broad possibilities in sensing technologies,
this EM band is of considerable importance.
Smaller scale metamaterial devices can also offer
more 
complex and interesting scenarios,
including 
tunable devices \cite{souk}, filters \cite{alek},
and nanoantennas \cite{liu}.
%
For larger scale ENZ metamaterials,
high directivity of an incident beam has been demonstrated \cite{enoch}.
This can be scaled down and extended to
composites containing an array of nanowires,
yielding a birefringent response with only one direction
possessing ENZ properties \cite{alek}.
A metamaterial grating
can be designed to also have properties akin
to ENZ media \cite{mocella}.

Often times, the structure being modeled is assumed isotropic.
Although this offers simplifications,
anisotropy is an inextricable feature of metamaterials
that  plays a crucial role in their EM response.
For instance, at optical and infrared frequencies,
incorporating anisotropy into a thin planar (nonmagnetic) waveguide
can result in behavior
indicative of double negative index  media \cite{pod}.
Anisotropic metamaterial 
structures can now be created that
contain elements that possess extreme electric
and magnetic responses to an incident beam. The 
inclusion of naturally anisotropic materials that are also
frequency dispersive (e.g., liquid crystals), 
allows
additional control in
beam direction.
It has also been shown that metamaterial structures requiring
anisotropic permittivity and permeability can be
created using tapered waveguides \cite{smol}.
By assimilating anisotropic
metamaterial leaky wave structures
within conventional radiative systems, 
the possibility exists 
to further control the emission characteristics.

Prompted by submicron experimental
developments, and potential 
beam manipulation 
involving metamaterials 
with vanishing dielectric response, 
we investigate
a planar anisotropic
system with an ENZ response
at near-ir frequencies along a given (longitudinal) direction. By ``freezing"
the phase in the longitudinal direction and tuning the electric and
magnetic responses in the transverse directions,
we will demonstrate
the ability to achieve remarkable emission control and directivity.
When excited by a source,
the direction of energy flow can
be due to
the propagation of localized
surface waves.
There can also exist
leaky waves, 
whereby the energy radiatively ``leaks" from the structure while 
attenuating longitudinally.
Indeed, there can be a complex 
interplay between the different type of allowed modes
whether radiated or guided, or
some other mechanism involving material absorption.
Through a judicious choice of parameters,
the admitted modes for the metamaterial can result
in radiation launched within
a narrow cone spanned by the
outflow of energy flux.

Some of the earliest works involving conventional 
leaky wave
systems reported
narrow beamwidth antennas with prescribed radiation angles \cite{tamir}, 
and
forward/backward leaky wave propagation in planar multilayered 
structures \cite{tamir2}.
In the microwave regime, photonic
crystals \cite{colak,micco,laroche} and transmission lines
can also can serve as leaky wave antennas \cite{lim}. 
More recently, a leaky wave metamaterial antenna exhibited 
broad side scanning at a single frequency \cite{lim}. 
The leaky wave 
characteristics have also been studied for grounded
single and double negative metamaterial slabs \cite{bacc}.
Directive emission in 
the microwave regime was demonstrated 
for magnetic metamaterials in which one of the components of $\bm \mu$ is small \cite{yuan}.
Nonmagnetic slabs 
can also yield varied beam directivity \cite{slab2}.

To begin our investigation, a harmonic time dependence $\exp(-i \omega t)$ for the TM fields is assumed.
The  planar structure
contains  a central biaxial anisotropic metamaterial of width $2d$ sandwiched between
the bulk superstrate and substrate, each of which
can
in general be anisotropic. 
The 
material in each region 
is assumed linear  with a  biaxial permittivity tensor, 
$\bm \epsilon_i = \epsilon_i^{xx} \hat{\bf x}\hat{\bf x} +\epsilon_i^{yy} \hat{\bf y}\hat{\bf y}+\epsilon_i^{zz} \hat{\bf z}\hat{\bf z}$.
Similarly, the biaxial magnetic response is represented
via ${\bm \mu}_i = \mu_i^{xx} \hat{\bf x}\hat{\bf x} +\mu_i^{yy} \hat{\bf y}\hat{\bf y}+\mu_i^{zz} \hat{\bf z}\hat{\bf z}$. 
The translational invariance in the $y$ and $z$ directions allows
the magnetic field  in the  $i$th layer,
${\bf H}_i$, to be written ${\bf H}_i = \hat{\bm y} h_i^y(x) e^{i(\gamma z - \omega t)}$,
and the electric field as  ${\bf E}_i = [\hat{\bm x} e_i ^x(x) + \hat{\bm z} e_i^z(x)] e^{i(\gamma z - \omega t)}$.
Here, $\gamma\equiv \beta +  i \alpha$ is the complex longitudinal propagation constant.
We focus on wave propagation occurring
in the positive $x$ direction, and
nonnegative $\beta$ and $\alpha$.
Upon matching the tangential $\bm E$ and $\bm H$ fields
at the boundary, we arrive at
the general dispersion equation that governs the allowed modes for this structure,
\begin{align} 
\label{disp1}
\epsilon^{zz}_2 k_{\perp,2} & (\epsilon^{zz}_3 k_{\perp,1}  +\epsilon^{zz}_1 k_{\perp,3})
+ \\ \nonumber 
&[(\epsilon^{zz}_2)^2 k_{\perp,1} k_{\perp,3}-\epsilon^{zz}_1 \epsilon^{zz}_3 k^2_{\perp,2}]
\tan(2 d k_{\perp,2})=0,
\end{align}
where the transverse wavevector in the superstrate (referred to as region 1), $k_{\perp,1}$,  is, 
\begin{align}
\label{k1}
k_{\perp,1} = \pm \sqrt{\epsilon_{1}^{zz}/\epsilon_1^{xx}(\beta^2-\alpha^2)-
k_0^2 \mu_1^{yy} \epsilon_1^{zz} + 2i\alpha \beta \epsilon_1^{zz}/\epsilon_1^{xx}}.
\end{align}
For the metamaterial region (region 2), we write
$k_{\perp,2}=\pm \sqrt{k_0^2 \mu_2^{yy} \epsilon_2^{zz} -
\gamma^2 \epsilon_{2}^{zz}/\epsilon_2^{xx}} $,
and for the substrate (region 3),
$k_{\perp,3} = 
\pm\sqrt{\gamma^2 \epsilon_{3}^{zz}/\epsilon_3^{xx}-
k_0^2 \mu_3^{yy} \epsilon_3^{zz} }$. 
The choice of sign in regions 1 and 3 plays an important role
in the determination of the physical nature 
of the type of mode solution that will
arise. The two roots associated with $k_{\perp,2}$,
results
in the same solutions to
Eq.~(\ref{disp1}).
The dispersion (Eq.~(\ref{disp1})) is also obtained from the poles
of the reflection coefficient for a plane wave incident from above on the structure.
The  transverse components of the ${\bm E}$  
field in region 1 are,
$e_1^z = - i k_{\perp,1}/(k_0 \epsilon^{zz}_1) H_1 e^{-k_{\perp,1} (x-d)}$,
$e_1^x = \gamma /(k_0 \epsilon^{xx}_1) h_1^y$,
and $h_1^y = H_1 e^{-k_{\perp,1} (x-d)}$, where $H_1$ is a
constant coefficient. 

Next, 
to disentangle  
the evanescent and leaky wave fields,
we
separate the wavevector $k_{\perp,1}$ into
its real and imaginary parts:
$k_{\perp,1} = \pm (q^- +  i q^+)$,
with $q^+$ and $q^-$ real.
The $k_{\perp,1}$,
$q^-$, and $q^+$ are in general related, depending
on $\sgn(\epsilon_1^{zz}\alpha \beta /\epsilon_1^{xx})$.
For  upward wave propagation ($+x$ direction), clearly 
we have $q^+ q^- \ge 0$.
It is also apparent that the parameter $q^-$
represents the inverse length scale
of wave increase along the transverse $x$-direction.
We are mainly concerned with the 
$k_{\perp,1}$ that
correspond to exponential wave increase
in the transverse direction while decaying  in $z$,
a hallmark of leaky waves. 
Although
leaky wave modes are not localized, they
can be excited by a point or line source which gives rise
to limited regions of space of EM wave amplitude increase
before eventually decaying.
When explicitly decomposing $k_{\perp,1}$ into its real and imaginary parts,
there is an intricate 
interdependence among
$\gamma$, $\bm \epsilon_i$, and $\bm \mu_i$ (for $\alpha \neq 0$):
$ q^{\pm}=1/\sqrt{2} (\sqrt{{\cal A}^2 + {\cal B}^2} \mp {\cal A})^{1/2}$,
where ${\cal A} = \epsilon_1^{zz}/\epsilon_1^{xx} (\beta^2-\alpha^2) - k_0^2 \mu_1^{yy} \epsilon_1^{zz}$,
and ${\cal B} = 2 \alpha \beta \epsilon_1^{zz}/\epsilon_1^{xx}$.
We will see below that $q^+$ is the root of interest in determining leaky wave emission for our structure.
At this point the surrounding media can have frequency dispersion in $\bm \epsilon_i$, 
and $\bm \mu_i$, while
the 
anisotropic metamaterial region can be dispersive and absorptive.


We are ultimately interested in anisotropic metamaterials
with an ENZ response along the axial direction ($z$-axis).
In the limit of vanishing $\epsilon_2^{zz}$, and perfectly conducting ground plane,
Eq.~(\ref{disp1}) can be solved analytically for the complex propagation constant, $\gamma$. The result is
\begin{widetext}
\begin{equation}
\label{gam}
\gamma^{\pm} = \dfrac{1}{\sqrt{2}}\dfrac{\sqrt{(\epsilon_{2}^{xx})^2+8(k_0 d)^2\epsilon_1^{xx}\epsilon_1^{zz}\epsilon_2^{xx}\mu_2^{yy}\pm 
\left|\epsilon_2^{xx}\right|
\sqrt{(\epsilon_{2}^{xx})^2+(4 k_0 d)^2\epsilon_1^{zz}\epsilon_1^{xx} (\mu_2^{yy}\epsilon_2^{xx}-\mu_1^{yy} \epsilon_1^{xx})}}}
{2 k_0 d\sqrt{\epsilon_1^{xx} \epsilon_1^{zz}}}.
\end{equation}
\end{widetext}
The two possible 
roots correspond to
distinct dispersion branches (seen below).
There are, in all, four solutions, $\gamma^{\pm}$,
and $-\gamma^{\pm}$.
The 
geometrical and material dependence
contained in Eq.~(\ref{gam}), 
determines the entire spectrum of the leaky
radiation fields
that may exist in our system.

There are numerous quantities one can study in order to effectively 
characterize leaky wave emission. One physically meaningful
quantity is the energy transport velocity, ${\bm v}_T$, which is 
the velocity at
which EM energy is transported through a medium \cite{loudon,ruppin}.
It is intuitively expressed as the ratio of the time-averaged
Poynting vector, ${\bm S}_{\rm avg}$, to the energy density, $U$:
${\bm v}_T \equiv {\bm S}_{\rm avg}/U$. 
Properly accounting for frequency dispersion that
may be present, we  can thus write the energy velocity for EM radiation emitted
above the structure,
\begin{align}
\label{vt}
{\bm v}_T = \dfrac{c/(8\pi){\rm Re}[ {\bm E}_1 \times {\bm H}_1^* ]}
{1/(16\pi) \Bigl[{\bm E}_1^\dagger \cdot 
\dfrac{d(\omega {\bm \epsilon}_1)}{d\omega} {\bm E}_1+
{\bm H}_1^\dagger \cdot
\dfrac{d(\omega {\bm \mu}_1)}{d\omega} {\bm H}_1
\Bigr]},
\end{align}
where  the conventional
definition \cite{shitz} of $U$ has been extended 
to include anisotropy.
Inserting the calculated EM fields, 
we get the compact expression (assuming no dispersion in the superstrate),
\begin{align}
\label{vt2}
{\bm v}_T = \omega \dfrac{\bigl( \epsilon_1^{xx} q^+ \hat{\bm x} + \epsilon_1^{zz} \beta \hat{\bm z}  \bigr)}
{\epsilon_1^{zz} \beta^2 + \epsilon_1^{xx} (q^+)^2}.
\end{align}
The corresponding direction of energy outflow is straightforwardly extracted from
the vector directionality in Eq.~(\ref{vt2}), 
\begin{equation}
\label{vt3}
\varphi =\tan^{-1}\bigl(\dfrac{\epsilon_1^{xx} q^+}{\epsilon_1^{zz} \beta} \bigr),
\end{equation}
which holds in the case of loss and frequency dispersion in the metamaterial.
It is evident that Eq.~(\ref{vt3}) satisfies 
$\varphi \rightarrow 0$ as $\alpha \rightarrow 0$,  
corresponding to the disappearance of
the radiation cone and possible emergence of guided waves.
In this limit, ${\bm v}_T = \hat{\bm z} \omega/\beta$, 
which corresponds to
the expected phase velocity, or velocity at which plane wavefronts travel
along the $z$-direction.
There is also angular symmetry, where 
$\varphi(\epsilon_2^{xx}) \rightarrow \varphi(-\epsilon_2^{xx})$, when $\mu_2^{yy} \rightarrow -\mu_2^{yy}$. 
For high refractive index media ($\epsilon_2^{xx}$ or $\mu_2^{yy}$ $\rightarrow \infty$),
we moreover recover the expected result
that $\varphi$ tends toward broadside ($\varphi = 0$).

Leaky wave emission from
an anisotropic metamaterial in vacuum 
\begin{figure}
\centering
\includegraphics[width=.483\textwidth]{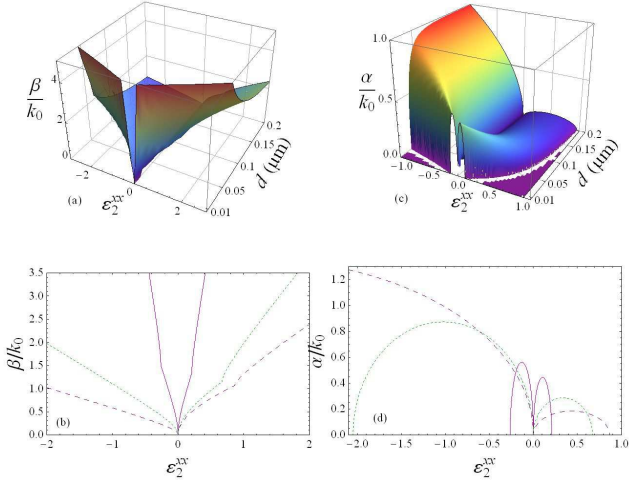}
\caption{(Color online). 
The real ($\beta$) and imaginary ($\alpha$) parts of the 
complex propagation constant $\gamma^+$, normalized by the vacuum wavevector $k_0$
at $f=280$ THz ($\mu_2^{yy} =1$). 
The figures (a) and (c)
are 3D global views depicting $\alpha$ and $\beta$ as functions of $\epsilon_2^{xx}$
and the thickness parameter $d$.
Figures (b) and (d) represent the normalized $\beta$ and $\alpha$, respectively, as
functions of $\epsilon_2^{xx}$ and for 
$d= 0.01$ \textmu m (solid curve), $d=0.05$  \textmu m (dotted curve), and $d=0.1$ \textmu m (dashed curve). 
}
\label{3D} 
\end{figure}
is characterized in Figs.~\ref{3D} (a) and (c), 
where 3-D views of the normalized $\beta$ (${\rm Re}[\gamma^+]$),
and $\alpha$  (${\rm Im} [\gamma^+]$),
are shown as functions
of the transverse dielectric response, $\epsilon_2^{xx}$, and thickness
parameter, $d$ (the width = $2d$). In Fig.~\ref{3D} (b) and (d), 2D slices depict the normalized
$\beta$ and $\alpha$ as functions of $\epsilon_2^{xx}$. 
Only the positive root, $\gamma^+$, is shown,
corresponding to the leaky wave case of interest, with $\alpha \ge 0$.
The slight kinks in the curves (see Fig.~\ref{3D}(b))
are 
at points where the $\gamma^-$
solutions would emerge (for $\alpha < 0$).
Both panels on the left clearly demonstrate
how $\beta$ rises
considerably with increasing $|\epsilon_2^{xx}|$. 
For subwavelength widths ($k_0 d\ll 1$), and to lowest order,
the propagation constant varies
linearly in $\epsilon_2^{xx}$, as $\beta/k_0 \approx \epsilon_2^{xx}/(2 k_0 d \sqrt{\epsilon_1^{xx} \epsilon_1^{zz}})$.
As the dielectric response $\epsilon_2^{xx}$ vanishes, corresponding
to an isotropic ENZ slab, we see from the figures (and Eq.~(\ref{gam}))
that $\beta \rightarrow 0$ (long wavelength limit), and emission
is subsequently perpendicular to the interface (see below).
It is also interesting that the important parameter $\alpha$ characterizing
leaky waves rapidly increases from zero at $\epsilon_2^{xx}=0 $ and peaks at differing values, 
depending
on the width of the emitting structure (Figs.~\ref{3D} (c) and (d)), 
until eventually returning to zero at the two
points, $\epsilon_2^{xx} =
4[-2 (k_0d)^2 \pm \sqrt{(k_0 d)^2 + 4 (k_0 d)^4}]$. 
This illustrates that $\alpha$ is spread over a greater
range of $\epsilon_2^{xx}$ for larger widths, but as previously discussed in conjunction with Fig.~\ref{3D},
$\alpha$ simultaneously suffers a dramatic reduction.
For small $d/\lambda$, the extremum of Eq.~(\ref{gam}), reveals that
the strength of the $\alpha$ peaks, $\alpha_{\rm max}$,
are given by $\alpha_{\rm max} \approx 1/2 \pm k_0 d$.

\begin{figure}
\centering
\includegraphics[width=.483\textwidth]{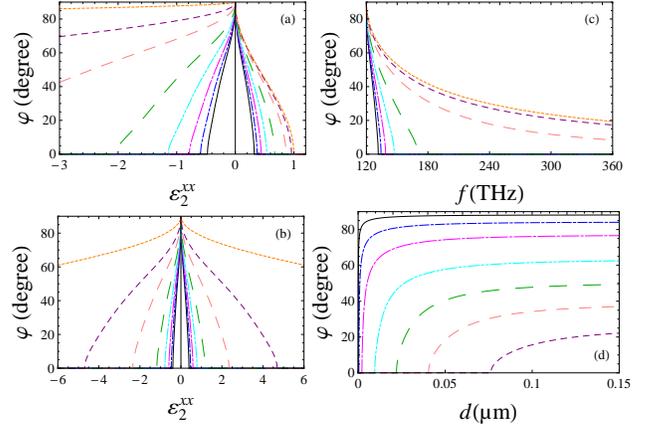}
\caption{(Color online) Leaky wave launch angle, $\varphi$, as a function
of permittivity, $\epsilon_2^{xx}$ (Figs.~(a) and (b))
for eight different
thicknesses in succession, 
starting with $d=1$ \textmu m (dotted orange curve), 
and subsequent values of $d$ (in \textmu m),
equaling $1/5,1/10,1/20,1/30,1/40,1/50$, and $1/60$.
Other parameters are as in Fig.~\ref{3D}.
In (c) the emission angle is shown as a function of frequency for the same
thicknesses in (a) and (b).
In (d), the effects of geometrical variation are presented  for 
$\epsilon_2^{xx} = 0.001,0.01,0.05,0.2,0.4,0.6$, and $0.8$. The curves with larger
overall $\varphi$ correspond to smaller $\epsilon_2^{xx}$ in succession.
}
\label{f3} 
\end{figure}

Next, in Fig.~\ref{f3},
we show the angle, $\varphi$, which defines the radiation cone from 
the surface of the metamaterial structure, as functions of both $\epsilon_2^{xx}$, 
frequency, and thickness parameter $d$. 
In panel (a),
the variation in $\varphi$ is shown over a
broad range of $\epsilon_2^{xx}$ for nonmagnetic media ($\mu_2^{yy} = 1$),
while panel (b) is for a metamaterial with vanishing $\mu_2^{yy}$,
representative of a type of matched impedance \cite{cuong}.
The eight curves in 
Fig.~\ref{f3} (a) and (b) 
represent different widths, identified in the caption. 
We see that for $\epsilon_2^{xx} \rightarrow 0$, we recover the
isotropic result of nearly normal emission ($\varphi \approx 90^{\rm o}$),
discussed and demonstrated in the millimeter regime \cite{enoch}.
This behavior can be understood in our system, at least qualitatively, from a geometrical
optics perspective and a generalization of Snell's Law for
bianisotropic media \cite{kong}.
When the magnetic response vanishes (Fig.~\ref{f3} (b)),
the emission angle becomes symmetric with respect to $\epsilon_2^{xx}$,
dropping from $\varphi = \pi/2$ for zero $\epsilon_2^{xx}$, to broadside ($\varphi =0$)
when $\epsilon_2^{xx} = \pm 4 k_0 d$. 
Thus thinner widths result in
more rapid beam variation as a function of $\epsilon_2^{xx}$.
In Fig.~\ref{f3}(c) we show how the emission angle varies  as a function of frequency,
with the transverse response obeying the Drude form,  
$\epsilon_2^{xx} =  1-\omega_p^2/(\omega^2 + i \Gamma \omega)$.
Here, $\omega_p = (2 \pi) 120$ THz and $\Gamma = 0$, to
isolate
leaky wave effects. 
With increasing frequency,
we observe similar trends found in the previous figures, where 
a larger dielectric response pulls 
the beam towards the metamaterial.
In (d), a geometrical study
illustrates how the emission angle varies with thickness: 
for $\epsilon_2^{xx} \mu_2^{yy} < 1$, the emission angle
rises abruptly with increased $d$, before leveling off at $\phi =
\tan^{-1}(\sqrt{1/(\epsilon_2^{xx} \mu_2^{yy}) -1})$. 
Physically, as the slab increases in size,
the complex propagation constant 
becomes purely real,  $\gamma \rightarrow \sqrt{\epsilon_2^{xx} \mu_2^{yy}}$,
and $q^+ \rightarrow \sqrt{1-\epsilon_2^{xx} \mu_2^{yy}}$.
This is consistent with what was discussed previously involving the depletion of $\alpha$ with $d$;
for thick ENZ slabs, leaky wave radiation is replaced by conventional propagating modes.
For fixed $\epsilon_2^{xx}$, 
there is also a critical thickness, $d^*$, below which no leaky waves are emitted,
which by Eq.~(\ref{gam}) is,  $d^* = \epsilon_2^{xx}/(4 k_0 \sqrt{1-\epsilon_2^{xx} \mu_2^{yy}})$.
\begin{figure}
\begin{center}
\subfigure{
\includegraphics[width=.23\textwidth]{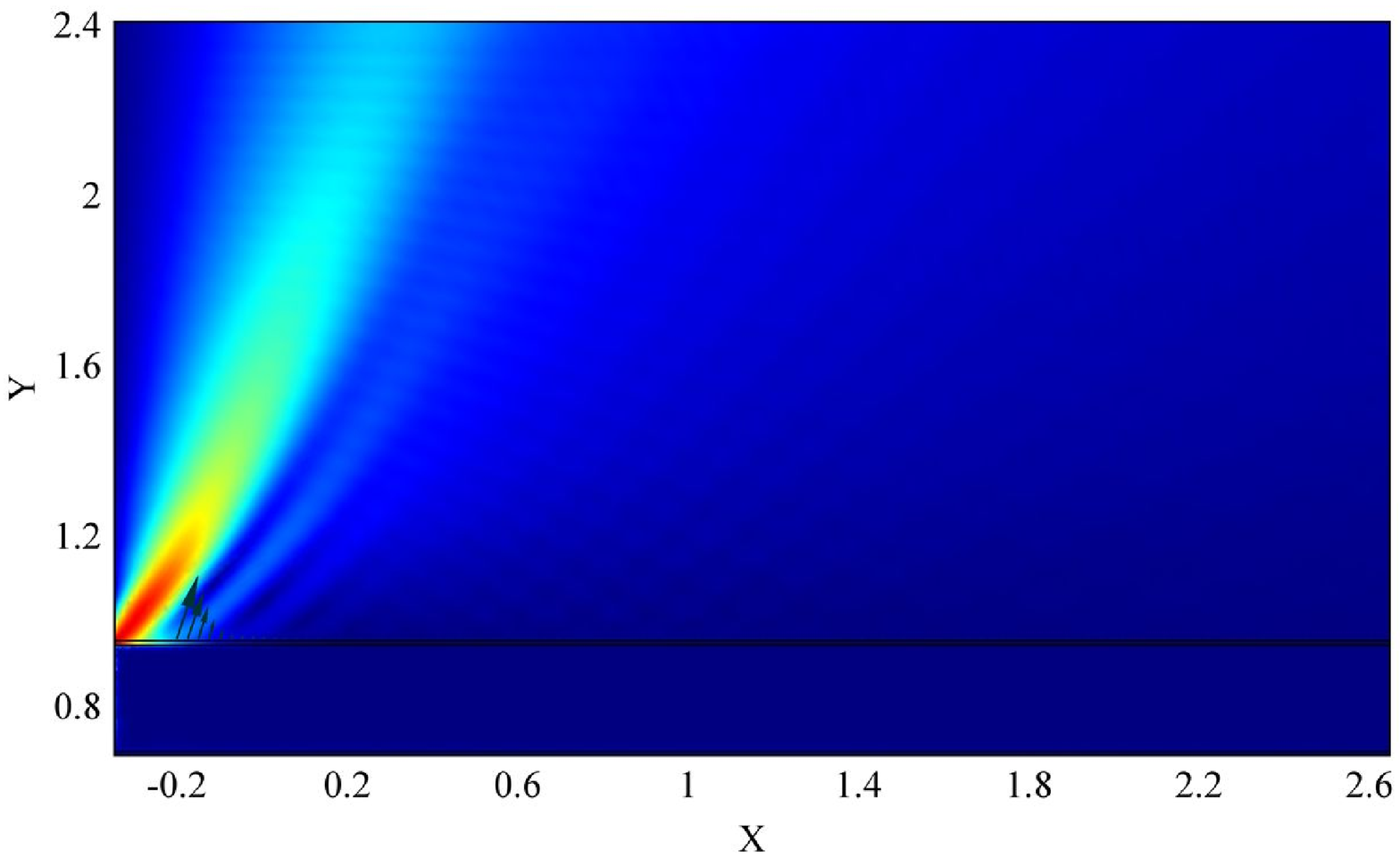}}
\subfigure{
\includegraphics[width=.23\textwidth]{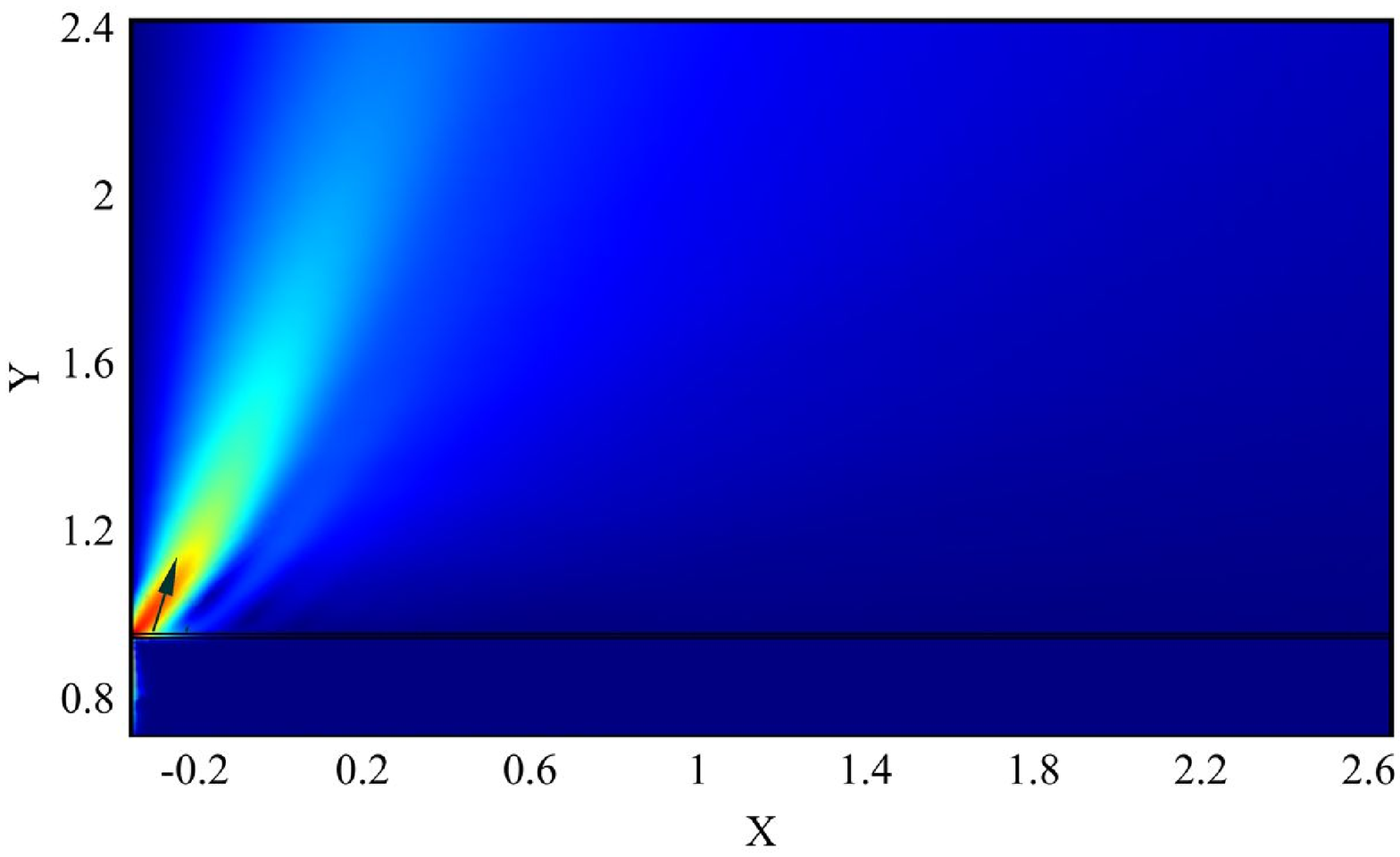}}
\subfigure{
\includegraphics[width=.23\textwidth]{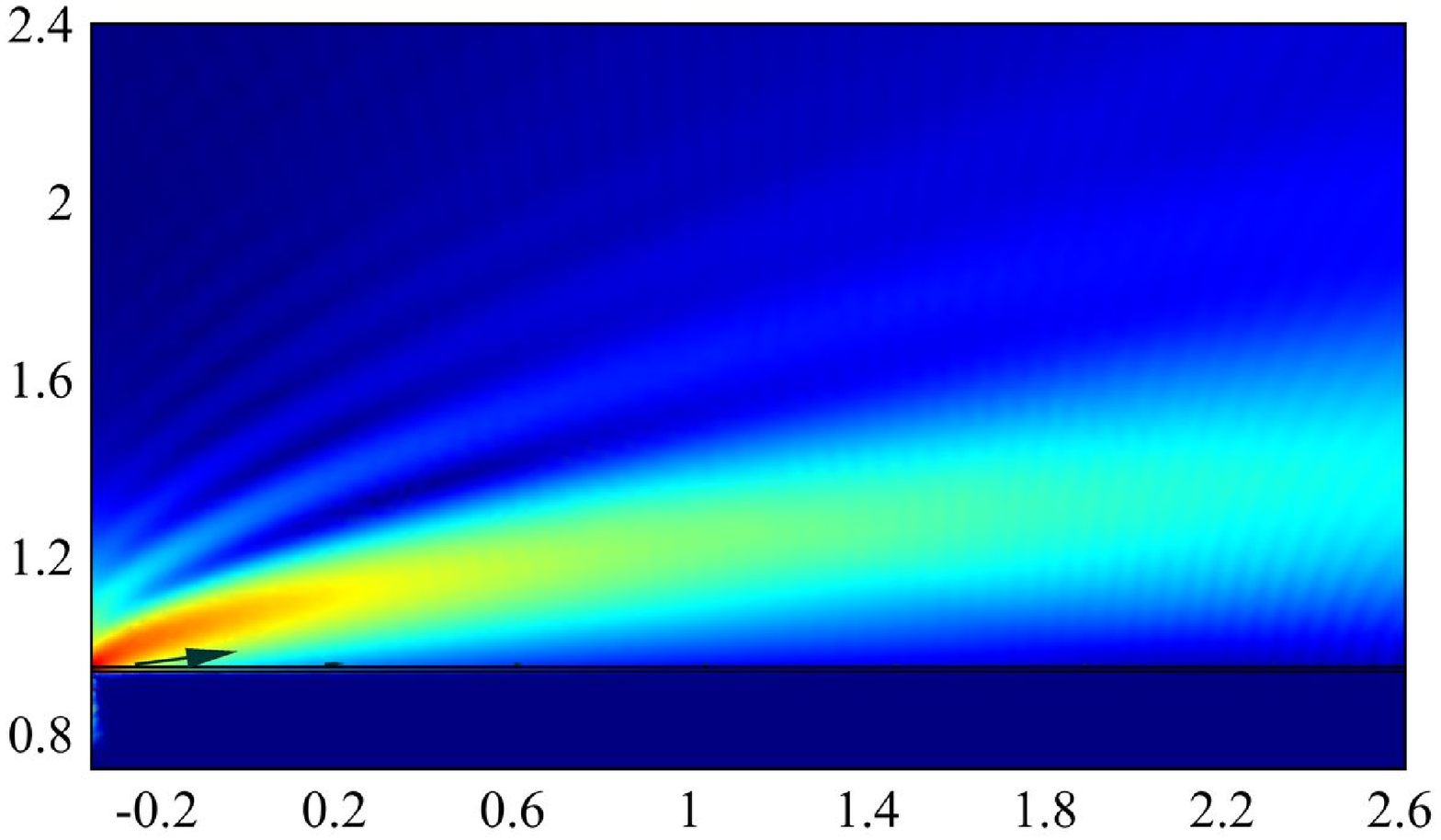}}
\subfigure{
\includegraphics[width=.23\textwidth]{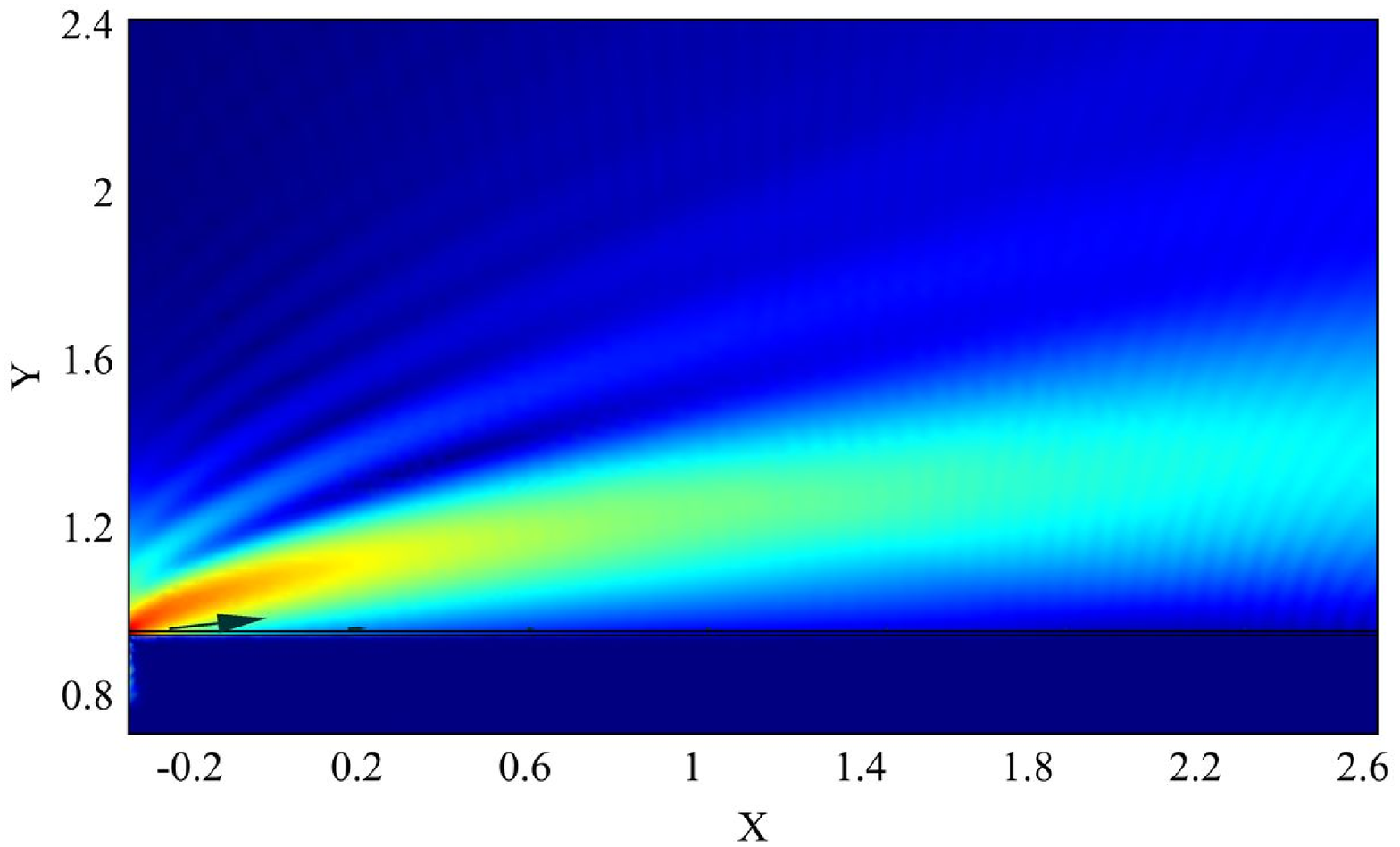}}
\caption{\label{cuong} Normalized field profiles
illustrating broad angular variation in beam emission.
The arrows along the interfaces depict the Poynting vector. 
The top left and right panels correspond to  $\epsilon_2^{xx} = 0.05$, and 
$\epsilon_2^{xx} = 0.05 + 0.02 i$, respectively. The
the bottom left and right panels are for $\epsilon_2^{xx} =0.66$  and $\epsilon_2^{xx} =0.66 + 0.02 i$, respectively. 
The metamaterial is subwavelength ($d=1/20$ \textmu m)
and nonmagnetic ($\mu_2^{yy} =1$). Coordinates are given in units of
$(\times 10)$ \textmu m.
}
\end{center}
\end{figure}
These results are consistent 
with simulations from a commercial finite element
software package \cite{comsol}. 
In Fig.~\ref{cuong}, we show the normalized $|{\bm H}|$ 
arising from a source excitation (at $f=280$ THz) within the metamaterial 
for $d=1/20$ \textmu m. 
The left two panels are for $\Gamma = 0$, and the right two have 
absorption present.
The full wave simulations agree with Fig.~\ref{f3}(a) (dashed green curve), 
where the leaky-wave energy outflow spans a broad angular range when  
$0\lesssim \epsilon_2^{xx} \lesssim 0.68$.
The right two panels exemplify the robustness of this effect 
for moderate amounts of loss present in the metamaterial.

In conclusion, we have demonstrated
leaky wave radiation in subwavelength 
biaxial metamaterials with vanishing permittivity
along the longitudinal direction.
The
leaky-wave radiation cone 
illustrated broad directionality through
variations in the transverse EM response. 
By utilizing nanodeposition techniques,
such  structures can be fabricated
by implementing an array of metallic nanowires
embedded in a self-organized porous nanostructured material \cite{alek}.

%
%


\acknowledgments
K.H. is supported in part by ONR 
and a grant 
of HPC resources as part of the DOD HPCMP.

\end{document}